\documentclass{elsart}

\usepackage{amssymb}
\usepackage{lscape}

\begin{document}

\begin{frontmatter}



\title{The empirical equilibrium structure of diacetylene\thanksref{birthday}}
\thanks[birthday]{This work is dedicated to Drs. Herbert M. Pickett and Edward A. Cohen  
in recognition of their many valuable contributions to the field of laboratory molecular spectroscopy.}


\author[mpifr]{Sven Thorwirth}
\ead{sthorwirth@mpifr-bonn.mpg.de}
\author[mainz]{Michael E. Harding}
\author[mpifr]{Dirk Muders}
\author[mainz]{J\"urgen Gauss}
\address[mpifr]{
Max-Planck-Institut f\"ur Radioastronomie, Auf dem H\"ugel 69, 53121 Bonn, Germany
}
\address[mainz]{Institut f\"ur Physikalische Chemie, Universit\"at Mainz, 55099 Mainz, Germany
}

\begin{abstract}
High-level quantum-chemical calculations are reported 
at the  MP2 and CCSD(T) levels of theory
for the equilibrium structure and the harmonic 
and anharmonic force fields of diacetylene, $\rm H-C\equiv C-C\equiv C-H$. 
The calculations were performed 
employing
Dunning's hierarchy of correlation-consistent basis
sets cc-pV$X$Z, cc-pCV$X$Z, and cc-pwCV$X$Z, as well as
the ANO2 basis set of Alml\"of and Taylor.
An empirical equilibrium structure based on 
experimental rotational constants for thirteen
isotopic species of diacetylene and computed zero-point
vibrational corrections is determined  ($r_e^{emp}$: $r_{\rm C-H}=1.0615$\,\AA, $r_{\rm C\equiv C}=1.2085$\,\AA, $r_{\rm C-C}=1.3727$\,\AA)
and in good agreement with the best theoretical 
structure (CCSD(T)/cc-pCV5Z: $r_{\rm C-H}=1.0617$\,\AA, $r_{\rm C\equiv C}=1.2083$\,\AA, $r_{\rm C-C}=1.3737$\,\AA ). In addition,
the computed fundamental vibrational frequencies are compared 
with the available experimental data and found
in satisfactory agreement.
\end{abstract}

\begin{keyword}
Diacetylene \sep Butadiyne \sep Rotation-vibration interaction \sep Anharmonic force field \sep Equilibrium structure
\end{keyword}
\end{frontmatter}

\section{Introduction}

Diacetylene (DiA), H$-$C$\equiv$C$-$C$\equiv$C$-$H, the simplest polyacetylene, is of importance in organic chemical synthesis \cite{maretina_RussChemRev_69_591_2000} and is also an abundant astronomical species. In space, it was detected through infrared spectroscopy in the atmosphere of Saturn's moon Titan \cite{kunde_nature_292_686_1981} and in the protoplanetary nebula CRL618 \cite{cernicharo_ApJ_546_L123_2001}.
In the laboratory, it has been the subject of numerous spectroscopic studies in the infrared and also in the microwave regime (see Refs. \cite{tay_StructChem_6_47_1995,matsumura_JMolSpectrosc_240_120_2006} and references therein). Spectroscopic data were also used to evaluate its structural parameters using different approaches.
An advanced experimental near-equilibrium $(r^{\rho}_m)_{corr}$ structure was reported by Tay et al. \cite{tay_StructChem_6_47_1995} making use of experimental rotation-vibration data of a total
of nine different isotopic species. Additionally, various quantum-chemical calculations
were performed with an empirically corrected frozen-core CCSD(T)/cc-pVQZ calculation \cite{botschwina_JMolSpectrosc_208_292_2001} representing the
highest theoretical level employed to date.


The equilibrium structure of DiA is 
of interest, as it represents one of the
simplest systems with conjugated carbon-carbon multiple
bonds. The question is how conjugation
affects the structure and in particular
shortens the central carbon-carbon single bond. 
It is well known that low-level quantum-chemical 
calculations are not reliable
in this respect, with Hartree-Fock calculations
typically underestimating the conjugation and 
density-functional theory overestimating it. 
High-level calculations, preferably
at the coupled-cluster (CC) level \cite{gauss_cc}, are needed 
for reliable predictions. To determine the accuracy 
of the CC calculations, it is essential to provide an 
equilibrium structure based on experimental data.

In a recent Fourier transform microwave spectroscopic study, the rotational
spectra of four new isotopologs of monodeutero diacetylene were characterized \cite{matsumura_JMolSpectrosc_240_120_2006} raising the number of known isotopic species
of DiA to a total of thirteen.
In the present work, we report an empirical (or semi-experimental)
equilibrium structure for DiA together with corresponding
high-level CC results. The key to the empirical structure is 
the calculation of accurate anharmonic force fields which enables
the determination of individual zero-point vibrational contributions

\begin{equation}
\Delta B_0^{calc}=\sum_{i}\alpha_i(d_i/2) 
\end{equation}

for each isotopolog. In Eq.~(1), the vibrational correction $\Delta B_0^{calc}$  is given as the sum of calculated rotation-vibration interaction constants
$\alpha_i$ with 
$d_i$ as the degeneracy being 1 for non-degenerate (stretching) vibrations and 2 for doubly degenerate (bending) vibrations.  
With those corrections, it is possible
to obtain empirical equilibrium values for the rotational constants
according to 

\begin{equation}
B_e^{emp}=B_0^{exp}+\Delta B_0^{calc}.
\end{equation}


The equilibrium structure is then obtained through a least-squares fit of 
the structural parameters
to the estimated equilibrium moments of inertia $I_e$ (see, e.g., Ref. \cite
{stanton_IntRevPhysChem_19_61_2000}).
In case of DiA these are the three bond distances $r_{\rm C-H}$, $r_{\rm C\equiv C}$, and $r_{\rm C-C}$.


\section{Computational Details}


Quantum-chemical calculations were performed using second-order 
M{\o}ller-Plesset (MP2) perturbation theory as well as the
coupled-cluster singles and doubles (CCSD) approach augmented
by a perturbative treatment of triple excitations (CCSD(T))
\cite{raghavachari_chemphyslett_157_479_1989}. The calculations were carried out using basis sets
from Dunning's hierarchy of correlation-consistent bases \cite{dunning_JCP_90_1007_1989}.
To be more specific, the cc-pV$X$Z ($X$=D$-$6) sets \cite{dunning_JCP_90_1007_1989}
have been used in frozen-core
(fc) MP2 and CCSD(T) calculations, while the cc-pCV$X$Z ($X$=T$-$5) 
and cc-pwCV$X$Z ($X$=T,Q)
sets \cite{woon_JCP_103_4572_1995,peterson_JCP_117_10548_2002} have been chosen for the all-electron 
calculations. Additional calculations were also performed with 
atomic natural orbital (ANO) basis sets from Alml\"of and Taylor \cite{almlof_JCP_86_4070_1987}.
The latter basis set, ANO2\footnote{The ANO2 set consists 
of a (13s8p6d4f2g/5s4p3d2f1fg) contraction for C and a (8s6p4d3f/4s3p2d1f) 
contraction for H.}, was employed within the fc approximation.

Equilibrium geometries were optimized using analytic gradient
techniques \cite{watts_chemphyslett_200_1-2_1_1992}. Harmonic, cubic, and semidiagonal quartic
force fields were then obtained using analytic second-derivative
techniques \cite{ stanton_IntRevPhysChem_19_61_2000,gauss_chemphyslett_276_70_1997}. The required third and fourth
derivatives for the anharmonic force fields were determined by additional
numerical differentiation as described in Refs. \cite{stanton_IntRevPhysChem_19_61_2000,stanton_JCP_108_7190_1998}.
The vibration-rotation interaction constants $\alpha_i$ were calculated
using formulas given in Ref. \cite{mills_alphas} based on lowest-order rovibrational
perturbation theory. Recalculation of the cubic force field is avoided by 
transforming the computed force field for the main isotopolog to the 
corresponding normal coordinate representations of the other isotopologs.
Since this is not possible for the semidiagonal quartic force field, 
fundamental frequencies, again determined using lowest-order rovibrational
perturbation theory \cite{mills_alphas}, are only reported for the main isotopic 
species.

All calculations have been performed with the Mainz-Austin-Budapest version
of {\sc acesii} \cite{aces_II}; some of the expensive (larger) calculations were made
possible by using our recent parallel implementation of CC energy as well as
first and
second derivative calculations \cite{harding_JChemTheoryComput_4_64_2008}.

\section{Source of Experimental Data and Least-Squares Fit}

Experimental $B_0$ values of
thirteen isotopologues studied through rotation-vibration
\cite{tay_StructChem_6_47_1995,arie_JMolSpectrosc_155_195_1992}
 and rotational spectroscopy \cite{matsumura_JMolSpectrosc_240_120_2006} are
 collected in Table \ref{alpha}.
The least-squares fits for the structural parameters have been performed 
including the data for all thirteen isotopologues.
The fits have been performed with respect to the moments of inertia using
the same weight for all considered isotopic species.

\section{Results and Discussion}

Table \ref{alpha} summarizes in addition to the experimental rotational constants 
$B_0$ also the corresponding calculated
vibrational corrections as obtained at the (fc)MP2/cc-pVTZ, (fc)CCSD(T)/cc-pVTZ,
(fc)CCSD(T)/cc-pVQZ, CCSD(T)/cc-pCVQZ, and (fc)CCSD(T)/ANO2 levels of theory. 
The vibrational corrections $\Delta B_0$ have been also computed with
other basis sets (not given here explicitly). Analysis of all these calculations allows one to draw 
the following conclusions (numbers given in the following
with respect to the parent isotopic species): (a) the use of polarized split-valence 
basis sets (i.e., cc-pV$X$Z) in all-electron calculations leads to too
large corrections (e.g. 3.80 MHz for MP2/cc-pVTZ in comparison to 2.1 MHz 
for (fc)MP2/cc-pVTZ); (b) MP2 calculations overestimate the vibrational
corrections in comparison with CCSD(T) computations (e.g., 2.1 MHz for
(fc)MP2/cc-pVTZ and 0.91 MHz for (fc)CCSD(T)/cc-pVTZ); (c) the use of
quadruple-zeta basis sets is recommended, as the smaller cc-pVTZ and
cc-pCVTZ basis sets yield vibrational corrections with rather large
errors (e.g. 0.91 MHz for (fc)CCSD(T)/cc-pVTZ in comparison with 
1.92 MHz for (fc)CCSD(T)/cc-pVQZ and 2.22 MHz for (fc)CCSD(T)/cc-pCVTZ
in comparison with 1.84 MHz for (fc)CCSD(T)/cc-pCVQZ); (d) the use
of the ANO2 set yields similar results as the corresponding
cc-pVQZ calculations.

From this analysis we also conclude that our computed vibrational
corrections (at the CCSD(T)/cc-pCVQZ and (fc)CCSD(T)/cc-pVQZ levels) 
should have an accuracy of about 0.2 to 0.3 MHz. This accuracy
is sufficient for obtaining an accurate equilibrium structure
of DiA from the empirical $B_e$ rotational constants and, in turn, it is more than
adequate for a theoretical prediction of ground state rotational constants $B_0$.
For the latter, the errors in the theoretical determination of high-level $B_e$ values due
to remaining basis-set errors and still missing electron-correlation
contributions certainly are larger than the zero-point
vibrational correction $\Delta B_0^{calc}$ \cite{puzzarini_inprep}.

Table \ref{struct} gathers the new extensive set of structural parameters for DiA. There, we report
the different empirical structures of DiA as obtained from using the
vibrational corrections given in Table \ref{alpha} and the pure theoretical structures based
on high-level CC calculations using a variety of basis sets.
In addition, structural parameters from the literature 
are given \cite{tay_StructChem_6_47_1995,matsumura_JMolSpectrosc_240_120_2006,botschwina_JMolSpectrosc_208_292_2001}.

Comparing our five different empirical $r_e^{emp}$ structures, we note that 
values of 1.062 \AA, 1.208 \AA, and 1.373 \AA\ are obtained for the
C$-$H distance, the C$\equiv$C triple bond, and C--C single bond, respectively, 
irrespective of the used vibrational corrections. The differences 
in the vibrational corrections affects the bond lengths only in the
fourth decimal. Nevertheless, the changes when going from the 
cc-pVTZ to the cc-pVQZ or cc-pCVQZ corrections are not entirely
negligible and we consider the results obtained with the 
(fc)CCSD(T)/cc-pVQZ, CCSD(T)/cc-pCVQZ, and CCSD(T)/ANO2
corrections the most
reliable. The high quality of these three sets of equilibrium 
distances is also seen through the fact that the residuals in the
fits are as small as 6 kHz for the maximum and 0.7 kHz for the root mean-square
deviations. The statistical uncertainties in all of the
fits are small being 0.0004 \AA\ (CC single bond), 0.0003 \AA\ (CC triple bond), 
and 0.0001 \AA\  (CH bond) and hence in all cases below
0.001 \AA.

Comparison of the derived empirical $r_e$ structures with the pure
theoretical geometries reveals a good agreement with the results
obtained at our best quantum-chemical level (CCSD(T)/cc-pCV5Z). 
With 0.001 \AA\ and less the discrepancies are  in the expected 
range and probably due to the neglect of higher excitations
in the CC treatment \cite{heckert}. A closer analysis of the 
extensive set of theoretical data in Table \ref{struct} furthermore
demonstrates the importance of core correlation, as the (fc)CCSD(T) 
calculations yield too long bond distances (by about 0.0015 \AA\ for
the CH distances and 0.003 \AA\ for the CC distances). 
However, core-correlation
effects can be accurately treated using an additivity assumption, since the 
distances obtained via
\begin{equation}
r_e \simeq r_e{\rm (pV6Z,fc)}+r_e{\rm (pwCVQZ,ae)}-r_e{\rm (pwCVQZ,fc)}
\end{equation}
are $r_{\rm C-H}=1.0618$\,\AA, $r_{\rm C\equiv C}=1.2083$\,\AA, and $r_{\rm C-C}=1.3737$\,\AA\  
and hence within $10^{-4}$\,\AA\ of the  CCSD(T)/cc-pCV5Z results.

For the sake of completeness, Table \ref{struct} also
gives parameters from other equilibrium or near-equilibrium 
structure determinations.
Reasonable agreement is found between the best empirical equilibrium structures
$r_e^{emp}$ ((fc) CCSD(T)/cc-pVQZ, CCSD(T)/cc-pCVQZ, and (fc) CCSD(T)/ANO2) and the near-equilibrium $(r^{\rho}_m)_{corr}$ structure of
Tay et al. \cite{tay_StructChem_6_47_1995} with the biggest discrepancy being in the
central C$-$C bond. Somewhat larger discrepancies are found in comparison against the
$r_e$ structure from Ref. \cite{matsumura_JMolSpectrosc_240_120_2006}.
However, the latter structure was only based on a rough empirical correction 
for the difference between the $r_s$ and $r_e$ structure of HC$_3$N (without determining
the vibrational corrections explicitly) 
and hence is not comparable from a rigorous point of view. Good agreement is found for the
structure given by Botschwina and Puzzarini \cite{botschwina_JMolSpectrosc_208_292_2001} where
a theoretical structure obtained at the CCSD(T)/cc-pVQZ level of theory was empirically corrected
through comparison against an empirical equilibrium structure of HC$_3$N.
Finally, a comparison
of the $r_e^{emp}$ structures against the $r_0$ structure in Table \ref{struct} reveals the significance of including vibrational 
corrections in structural determinations. In the latter structure the CH distances are significantly 
shorter (by about 0.005 \AA) whereas 
the CC single bond is longer (by about 0.002 \AA) than in the $r_e^{emp}$ structures. 
A similar conclusion also holds for the comparison with the $r_s$ structure reported in
Ref.~\cite{matsumura_JMolSpectrosc_240_120_2006} for HCCCCD.

Comparison of the $r_e^{emp}$ parameters of diacetylene with those of similar mole\-cules reveals
that the C--H bond lengths found in diacetylene (1.6016 \AA , this work), fluoroacetylene (1.6014 \AA, Ref. \cite{botschwina_JMolSpectrosc_208_292_2001}),
methyldiacetylene (1.6013 \AA, Ref. \cite{cazzoli_submitted}), acetylene
(1.0618 \AA , Ref. \cite{cazzoli_JMolSpectrosc_247_115_2008}), HC$_3$N (1.0623 \AA , Ref.
\cite{botschwina_MolPhys_103_1441_1460}) and methylacetylene (1.061 \AA , Ref. 
\cite{leguennec_JMolSpectrosc_160_471_1993})
are practically identical. 
Clearly, conjugation effects in DiA manifest themselves in the C$\equiv$C bond length: whereas in acetylene 
a value of 1.2029(1) \AA\ is found \cite{cazzoli_JMolSpectrosc_247_115_2008},
the C$\equiv$C triple bond is significantly longer in diacetylene, namely 1.2084 \AA . The latter value is in good agreement
with those found in similar conjugated systems. Recent examples are CC distances in the ethynyl group of 
the substituted diacetylenes HC$_4$F and CH$_3$C$_4$H (1.2080 \AA\ \cite{botschwina_JMolSpectrosc_208_292_2001}
and 1.2085 \AA\ \cite{cazzoli_submitted}, respectively)
and also the C$\equiv$C distances in
branched species such as {\it cis}-Hex-3-ene-1,5-diyne  and ({\it Z)}-pent-2-en-4-ynenitrile (1.208(3) \AA\ \cite{mcmahon_JACS_122_939_2000} and 1.207(3) \AA\ \cite{halter_JACS_123_12353_2001}).
Given the prototypical C$=$C 
equilibrium bond length in ethylene of 1.3305 \AA\ \cite{craig_JPCA_110_7461_2006})
the formal C$-$C single bond length in DiA of 1.3727 \AA\ found here is very short
(and comparable to monofluorodiacetylene (1.3729 \AA) and methyldiacetylene (1.3734 \AA) 
indicative of some double-bond character. 
Furthermore, conjugation i.e. $\pi$ electron delocalization is more pronounced in DiA (and HC$_4$F/CH$_3$C$_4$H) than in HC$_3$N where the C$\equiv$C bond is found to be shorter (intermediate between acetylene and DiA) and the C$-$C bond longer compared to DiA (Table 1).

Several infrared studies of the vibrational fundamentals, overtones and hot bands of DiA are found in the literature (see Ref. \cite{tay_StructChem_6_47_1995} and references therein).
In Table \ref{vib} we report harmonic and fundamental 
frequencies of DiA as obtained through CCSD(T) calculations of the cubic and semi-diagonal quartic force fields using the 
cc-pVTZ, cc-pVQZ, and ANO2 basis sets in the fc approximation
and the cc-pCVQZ basis correlating  all-electrons. The accuracy of the reported fundamental frequencies
obtained using Dunning's quadruple-zeta basis sets is probably only a couple 
of wavenumbers, as it turned out impossible to converge the required CC
second-derivative calculations at the displaced points with very tight convergence 
thresholds. 
In addition, we also report the corresponding
(fc)MP2/cc-pVTZ results.
Comparison of the theoretical anharmonic (fundamental) frequencies
against the experimental values collected in Ref. \cite{mcnaughton_JMolStruct_273_11_1992} reveals that the 
results obtained with the cc-pV$X$Z ($X$=T, Q) sets turn out less reliable, while better agreement with experiment
is seen for the other two basis sets. For both, the cc-pCVQZ and the ANO2 set, the computed frequencies match
the experimental values within about 10 cm$^{-1}$. This finding again documents the suitability of the 
ANO basis sets for frequency calculations, while the Dunning basis sets (at least when using the valence
sets) show some deficiencies \cite{freqano_1,freqano_2,freqano_3,freqano_4,freqano_5}.

\section{Conclusions}

The equilibrium structure of diacetylene has been determined based on
the combination of experimental rotational constants $B_0$ of thirteen isotopic species
and zero-point vibrational corrections $\Delta B_0$ calculated at various quantum-chemical
levels.
The empirical equilibrium structures obtained agree to within 10$^{-3}$\AA\ irrespective of the
theoretical level employed.
From the present study, the new recommended equilibrium structure of DiA is
$r_{\rm C-H}=1.0615$\,\AA, $r_{\rm C\equiv C}=1.2085$\,\AA,  and $r_{\rm C-C}=1.3727$\,\AA .
This structure is in good agreement with complementary high-level CC calculations performed here.
The structural parameters at the highest level of theory (CCSD(T)/cc-pCV5Z) are  
$r_{\rm C-H}=1.0617$\,\AA, $r_{\rm C\equiv C}=1.2083$\,\AA, $r_{\rm C-C}=1.3737$\,\AA.

Evaluation of cubic and semi-diagonal quartic force fields calculated at
the CCSD(T)/cc-pCVQZ and (fc)CCSD(T)/ANO2 levels yielded harmonic and anharmonic 
vibrational frequencies being in good agreement with experiment, while 
corresponding (fc)MP2 and (fc)CCSD(T) calculations with the cc-pV$X$Z ($X$=T,Q) sets
are less satisfactory.

\textbf{Acknowledgment}\\
This work at Mainz has been supported by the Deutsche Forschungsgemeinschaft
(DFG) and the Fonds der Chemischen Industrie.


\begin{landscape}
\begin{table*}[t]
\caption{Experimental rotational constants $B_0$ and computed zero-point
vibrational corrections $\Delta B_0$ (all values in MHz) for the
various isotopologues of diacetylene. \label{alpha}}
\begin{tabular*}{\hsize}{@{\extracolsep{\fill}}llrrrrr}
\hline
  &     &   \multicolumn{5}{c}{$\Delta B_0$}\\
\cline{3-7}
Isotopic  &   &  (fc)MP2/ & (fc)CCSD(T)/ & (fc)CCSD(T)/ & CCSD(T)/ & (fc)CCSD(T)/ \\
Species  & $B_0$ & cc-pVTZ & cc-pVTZ & cc-pVQZ & cc-pCVQZ & ANO2  \\
\hline
HCCCCH                             & 4389.3019(39)$^d$  &  2.098   &    0.912 &    1.922 &    1.838  &    2.083  \cr 
DCCCCD                             & 3809.2433(66)$^e$  & $-$0.207 & $-$1.110 & $-$0.203 & $-$0.237  & $-$0.005  \cr
H$^{13}$C$^{13}$C$^{13}$C$^{13}$CH & 4098.8959(36)$^e$  &  2.145   &    1.064 &    1.970 &    1.890  &    2.108  \cr
H$^{13}$C$^{13}$CCCH               & 4243.7325(111)$^e$ &  2.130   &    0.997 &    1.953 &    1.871  &    2.103    \cr
H$^{13}$CCCCH                      & 4258.5465(105)$^e$ &  2.112   &    0.953 &    1.912 &    1.829  &    2.063    \cr
HC$^{13}$CCCH                      & 4371.6291(45)$^e$  &  2.113   &    0.955 &    1.962 &    1.879  &    2.121    \cr
H$^{13}$C$^{13}$C$^{13}$CCH        & 4224.7392(99)$^e$  &  2.138   &    1.032 &    1.987 &    1.905  &    2.134    \cr
H$^{13}$C$^{13}$CC$^{13}$CH        & 4115.0556(42)$^e$  &  2.133   &    1.027 &    1.935 &    1.855  &    2.075    \cr
HCCCCD                             & 4084.45342(7)$^f$  &  0.815   & $-$0.219 &    0.745 &    0.688  &    0.927    \cr
H$^{13}$CCCCD                      & 3964.11797(17)$^f$ &  0.871   & $-$0.141 &    0.774 &    0.718  &    0.946    \cr
HC$^{13}$CCCD                      & 4066.49893(16)$^f$ &  0.830   & $-$0.178 &    0.782 &    0.725  &    0.962    \cr
HCC$^{13}$CCD                      & 4071.64202(16)$^f$ &  0.851   & $-$0.160 &    0.800 &    0.743  &    0.980    \cr
HCCC$^{13}$CD                      & 3977.69016(15)$^f$ &  0.870   & $-$0.144 &    0.775 &    0.718  &    0.947    \cr
\hline
\end{tabular*}
$^a$ Ref. (fc)CCSD(T)/cc-pVQZ  \\
$^c$ Ref. (ae)CCSD(T)/cc-pCVQZ \\
$^d$ Ref. \cite{arie_JMolSpectrosc_155_195_1992}.\\
$^e$ Ref. \cite{tay_StructChem_6_47_1995}.\\
$^f$ Ref. \cite{matsumura_JMolSpectrosc_240_120_2006}.
\end{table*}
\end{landscape}
\newpage

\begin{table*}[t]
\caption{Equilibrium structures of diacetylene and related molecules (\AA).\label{struct}}
\begin{tabular*}{\hsize}{@{\extracolsep{\fill}}lccc}
\hline
Method & $r_{\rm C-H}$    & $r_{\rm C\equiv C}$   & $r_{\rm C-C}$  \cr

\hline
(fc)MP2/cc-pVTZ              & 1.0620 & 1.2194 & 1.3687 \cr
(fc)CCSD(T)/cc-pVTZ          & 1.0638 & 1.2149 & 1.3789 \cr
(fc)CCSD(T)/cc-pVQZ          & 1.0633 & 1.2119 & 1.3769 \cr
(fc)CCSD(T)/ANO2             & 1.0631 & 1.2118 & 1.3766 \cr
(fc)CCSD(T)/cc-pVQZ+corr (Ref. \cite{botschwina_JMolSpectrosc_208_292_2001})$^a$  & 1.0615 & 1.2087 & 1.3720  \cr
(fc)CCSD(T)/cc-pV5Z          & 1.0630 & 1.2111 & 1.3764 \cr
(fc)CCSD(T)/cc-pV6Z          & 1.0630 & 1.2109 & 1.3762 \cr
CCSD(T)/cc-pCVTZ             & 1.0633 & 1.2121 & 1.3770 \cr
CCSD(T)/cc-pwCVTZ            & 1.0630 & 1.2111 & 1.3763  \cr
CCSD(T)/cc-pCVQZ             & 1.0621 & 1.2091 & 1.3742 \cr 
CCSD(T)/cc-pwCVQZ            & 1.0620 & 1.2089 & 1.3741  \cr
(fc)CCSD(T)/cc-pwCVQZ        & 1.0632 & 1.2115 & 1.3766  \cr
CCSD(T)/cc-pCV5Z             & 1.0617 & 1.2083 & 1.3737  \cr \hline
$r_0$                        & 1.0561 & 1.2079 & 1.3752 \cr
$(r^{\rho}_m)_{corr}$ (Ref. \cite{tay_StructChem_6_47_1995})& 1.0613(1) & 1.2096(1) & 1.3708(2)\cr
$r_e$ (Ref. \cite{matsumura_JMolSpectrosc_240_120_2006})   & 1.0609 & 1.2104 & 1.3709  \cr

$r_e^{emp}$ ((fc) MP2/cc-pVTZ)     & 1.0623 & 1.2077 & 1.3736  \cr
$r_e^{emp}$((fc) CCSD(T)/cc-pVTZ)  & 1.0620 & 1.2083 & 1.3732  \cr
$r_e^{emp}$((fc) CCSD(T)/cc-pVQZ)  & 1.0616 & 1.2084 & 1.3727 \cr 
$r_e^{emp}$(CCSD(T)/cc-pCVQZ) & 1.0615 & 1.2085 & 1.3727 \cr 
$r_e^{emp}$((fc) CCSD(T)/ANO2)     & 1.0614 & 1.2084 & 1.3726 \cr 
\hline
$r_e^{emp}$ Acetylene (Ref. \cite{cazzoli_JMolSpectrosc_247_115_2008}) &  1.0618 & 1.2029 & --- \cr
$r_e^{emp}$ HC$_3$N (Ref. \cite{botschwina_MolPhys_103_1441_1460})     &  1.0623 & 1.2059 & 1.3761 \cr
$r_e^{emp}$ HC$_4$F (Ref. \cite{botschwina_JMolSpectrosc_208_292_2001})$^b$ &  1.0614 & 1.2080 & 1.3731 \cr
$r_e^{emp}$ H$_3$C$-$C$_4-$H (Ref. \cite{cazzoli_submitted}) & 1.6013(3) & 1.2085(6)/1.2091(16) & 1.3734(14)
\cr
\hline
\end{tabular*}
$^a$ Empirically corrected {\it ab initio} structure, see text.\\
$^b$ $r_{\rm C\equiv C}$ given refers to the ethynyl group.\\
\end{table*}

\newpage

\begin{landscape}
\begin{table*}[t]
\caption{Vibrational fundamentals of diacetylene (in cm$^{-1}$).$â$\label{vib}}
\label{res:gss}
\begin{tabular*}{\hsize}{@{\extracolsep{\fill}}lccccccccccc}
\hline
& Experiment$^b$  &
\multicolumn{2}{c}{(fc)MP2/cc-pVTZ} & 
\multicolumn{2}{c}{(fc)CCSD(T)/cc-pVTZ} & 
\multicolumn{2}{c}{(fc)CCSD(T)/cc-pVQZ} & 
\multicolumn{2}{c}{CCSD(T)/cc-pCVQZ} &
\multicolumn{2}{c}{(fc)CCSD(T)/ANO2} \cr
&                 & harm.  & anharm. & 
harm. & anharm. &
harm. & anharm. &
harm. & anharm. &
harm. & anharm. \cr
\hline
$\nu_1(\Sigma_g^+)$   &  3332 & 3486 & 3361 & 3458 & 3329 & 3457 & 3281 &  3463 & 3333 & 3462 & 3330  \cr 
$\nu_2(\Sigma_g^+)$   &  2189 & 2194 & 2147 & 2233 & 2188 & 2235 & 2184 &  2243 & 2197 & 2237 & 2190 \cr 
$\nu_3(\Sigma_g^+)$   &   872 &  897 &  867 &  887 &  849 &  892 &  855 &   894 &  865 &  891 &  861 \cr 
$\nu_4(\Sigma_u^+)$   &  3334 & 3482 & 3369 & 3454 & 3339 & 3458 & 3288 &  3465 & 3338 & 3461 & 3329\cr 
$\nu_5(\Sigma_u^+)$   &  2022 & 2008 & 1970 & 2051 & 2014 & 2057 & 2016 &  2064 & 2031 & 2056 & 2020 \cr 
$\nu_6(\Pi_g)$        &   626 &  620 &  625 &  623 &  643 &  632 &  616 &   636 &  635 &  639 & 627  \cr 
$\nu_7(\Pi_g)$        &   483 &  487 &  452 &  474 &  419 &  481 &  484 &   484 &  490 &  482 & 479 \cr 
$\nu_8(\Pi_u)$        &   628 &  629 &  602 &  633 &  614 &  634 &  616 &   640 &  635 &  640 & 627 \cr 
$\nu_9(\Pi_u)$        &   220 &  230 &  217 &  227 &  216 &  220 &  225 &   221 &  223 & 220 &  219  \cr 
\hline
\end{tabular*}
The accuracy of the reported fundamental frequencies
obtained using Dunning's quadruple-zeta basis sets is probably only a couple 
of wavenumbers, as it turned out impossible to converge the required CC
second-derivative calculations at the displaced points with very tight convergence 
thresholds.
$^b$ Ref. \cite{mcnaughton_JMolStruct_273_11_1992} and references therein.\\
\end{table*}
\end{landscape}

\end{document}